\documentclass[twocolumn,secnumarabic,amssymb, amsmath, nofootinbib,tightenlines,
nobibnotes, aps, prl,epsfig]{revtex4}
\usepackage{graphicx}
\usepackage{dcolumn}
\usepackage{bm}
\begin{document}
\preprint{APS/123-QED}
\title{A determination of the longitudinal structure function $F_{L}$ from the parametrization of $F_{2}$ based on the  Laplace
transformation }

\author{G.R.Boroun}%
 \email{grboroun@gmail.com; boroun@razi.ac.ir }
\affiliation{Department of Physics, Razi University, Kermanshah
67149, Iran}
\date{\today}
\begin{abstract}
I calculate the longitudinal structure function, using Laplace
transform techniques, from the parametrization of the structure
function $F_{2}(x,Q^{2})$ and its derivative at low values of the
Bjorken variable $x$. I consider the effect of the charm quark
mass to the longitudinal structure function, which leads to
rescaling variable for $n_{f}=4$. The results are compared with
the H1 collaboration data and the CTEQ-Jefferson Lab (CJ)
collaboration [Phys.Rev.D {\bf93}, 114017 (2016)] parametrization
model. The obtained results with the Bjorken variable of $x$ are
found to be comparable with the results L.P.Kaptari et al.
[Phys.Rev.D {\bf99}, 096019 (2019)] which is based on the Mellin
transform techniques.\\
\end{abstract}
 \pacs{***}
\keywords{****} 
\maketitle
\subsection{Introduction}

Understanding the basic internal structure of nucleon and the
quest for the ultimate constituents has always been important in
high energy physics. In ultra-high energy processes, at extremely
high inelasticity $y$, the longitudinal structure function becomes
predominant and its behavior will be extended down to very low $x$
in the Large Hadron electron Collider (LHeC). The Large Hadron
Electron Collider (LHeC), a possible future upgrade of the LHC,
will extend the HERA DIS measurement into a much smaller region of
$x$ and larger region of $Q^{2}$. In this region the longitudinal
structure function becomes predominant. In the LHeC design report
[1], a simulation study at low $x$ for the longitudinal structure
function has been performed in considerable detail. This has
recently been updated [2], with still enlarged luminosity and
improved detector systematics. The $F_L$ measurements with the
LHeC will reach a precision of a few percent. They are considered
to be extended down to $x$ below $10^{-6}$ with the the Future
Circular Collider electron- hadron (FCC-eh). In the future, the
electron-proton colliders will generate even more data with lower
$x$ values and high values of $Q^{2}$. The longitudinal structure
function is directly related to the gluon distribution in the
proton and its behavior has been predicted by Altarelli and
Martinelli [3] equation. Indeed, this represents a crucial test on
the validity of the perturbative QCD (pQCD) framework at small
$x$, when compared the experimental data with theoretical
predictions.\\
The recent articles in Ref.[4] by Kaptari $et~ al$. revives the
parametrization of the longitudinal structure function by using
the parametrization of the structure function $F_{2}$. The
parametrization of the proton structure function [5] is a fit to
HERA data on deep-inelastic lepton-nucleon scattering (DIS) at low
$x$ in a wide range of the momentum transfer $Q^{2}$,
$(1~\mathrm{GeV}^{2}<Q^{2}<3000~\mathrm{GeV}^{2})$. This fit
describes fairly well the available experimental data on the
reduced cross sections, at asymptotically low $x$, and provides
the parametrization of the structure function $F_{2}$ which is
relevant in investigations of ultra-high energy processes.  This
parametrization [5] provides reliable structure function $F_{2}$
by the following form
\begin{eqnarray}
F_{ 2}(x,Q^{2})& =& D(Q^{2})(1-
x)^{n}\sum_{m=0}^{2}A_{m}(Q^{2})L^{m},
\end{eqnarray}
where  the explicit expression for the proton structure function
and  effective parameters are defined in Appendix A and Table I.
The authors of Ref.[4] show that the longitudinal structure
function in the leading-order (LO) approximation is considered in
the same form of the parametrization of the structure function
$F_{2}$ (i.e., Eq.(1)), which is
\begin{eqnarray}
F_{L}(x,Q^{2})&=&(1-x)^{\nu}\sum_{m=0}^{2}C_{m}(Q^{2})L^{m}.
\end{eqnarray}
The parametrization of structure functions (i.e., Eqs.(1) and
(2)), suggested in Refs.[4] and [5], are in a full accordance with
the Froissart predictions. In this paper I introduce a method to
calculate $F_{L}(x,Q^{2})$ inside the proton by using the Laplace
transform techniques. Several methods to relate $F_{L}$ and
$F_{2}$ scaling violation to the gluon density at small $x$ were
suggested previously [6-12]. To study the longitudinal structure
function I use the Laplace-transform technique for solving the
Altarelli- Martinelli equation by employing the parametrization of
the structure function $F_{2}$.\\

\subsection{Method}

In $s$-space the obtained equations [13-18] for the structure
functions, which relate both the quark and gluon densities,
defined by the following forms as
\begin{eqnarray}
\frac{\partial{f_{2}(s,Q^{2})}}{\partial{\ln}Q^{2}}&=&
\Phi_{f}(s)f_{2}(s,Q^{2})+<e^{2}>\Theta_{f}(s)g(s,Q^{2}),\nonumber\\
f_{L}(s,Q^{2})&=&\Phi_{L}(s)f_{2}(s,Q^{2})
+<e^{2}>\Theta_{L}(s)g(s,Q^{2}).\nonumber\\
\end{eqnarray}
The Laplace-transform of the distribution functions read
\begin{eqnarray}
{\mathcal{L}}[\mathcal{\widehat{F}}_{L}(\upsilon,Q^{2});s]&=&f_{L}(s,Q^{2}),\nonumber\\
{\mathcal{L}}[\mathcal{\widehat{F}}_{2}(\upsilon,Q^{2});s]&=&f_{2}(s,Q^{2})
\end{eqnarray}
where
\begin{eqnarray}
\mathcal{\widehat{F}}_{L}(\upsilon,Q^{2})&=&\frac{4\pi}{\alpha_{s}(Q^{2})}F_{L}(e^{-\nu},Q^{2}),\nonumber\\
\frac{\partial{\mathcal{\widehat{F}}_{2}(\upsilon,Q^{2})}}{\partial{\ln}Q^{2}}&=&
\frac{4\pi}{\alpha_{s}(Q^{2})}\frac{{\partial}F_{2}(e^{-\nu},Q^{2})}{\partial{\ln}Q^{2}}.
\end{eqnarray}
The $\nu$ variable is $\nu{\equiv}\ln(1/x)$ and  $<e^{2}>$ is the
average of the charge $e^{2}$ for the active quark flavors,
$<e^{2}>=n_{f}^{-1}\sum_{i=1}^{n_{f}}e_{i}^{2}$. It should be
noted that the Laplace transform of convolution factors is simply
the
ordinary product of the Laplace transform of the factors.\\
The coefficient functions $\Phi$ and $\Theta$ in $s$-space are
given by
\begin{eqnarray}
\Phi_{L}(s)&=&4C_{F}\frac{1}{2+s},\nonumber\\
\Theta_{L}(s)&=&8n_{f}(\frac{1}{2+s}-\frac{1}{3+s}),\nonumber\\
\Theta_{f}(s)&=&2n_{f}(\frac{1}{1+s}-\frac{2}{2+s}+\frac{2}{3+s}),\nonumber\\
\Phi_{f}(s)&=&4-\frac{8}{3}(\frac{1}{1+s}+\frac{1}{2+s}+2(\psi(s+1)+\gamma_{E})),\nonumber\\
\end{eqnarray}
where $\psi(x)$ is the digamma function and
$\gamma_{E}=0.5772156..$ is Euler constant. For the SU(N) gauge
group, $C_{F}=4/3$ is the color Cassimir operator in QCD. In the
above equations, $\psi(s)$ is defined by
$\psi(s)=\frac{d}{ds}{\ln}\Gamma(s)$ and $S(s)=\psi(s+1)-\psi(1)$
expressed [4] by
$S(s)=-{\ln}2-\sum_{l=0}^{\infty}\frac{(-1)^{l+1}}{s+l+1}$.\\
Eventually, the longitudinal structure function is defined into
the proton structure function and the derivative of the proton
structure function with respect to $\ln{Q^{2}}$ in $s$-space by
the following form
\begin{eqnarray}
f_{L}(s,Q^{2})=\frac{\alpha_{s}(Q^{2})}{4\pi}k(s)f_{2}(s,Q^{2})+h(s)\frac{\partial{f_{2}(s,Q^{2})}}{\partial{\ln}Q^{2}},
\end{eqnarray}
where
$k(s)=\Phi_{L}(s)-\frac{\Theta_{L}(s)}{\Theta_{f}(s)}\Phi_{f}(s)$
and $h(s)=\frac{\Theta_{L}(s)}{\Theta_{f}(s)}$. The inverse
Laplace transform of $k(s)$ and $h(s)$ is given by the kernels
$\widehat{\eta}(\nu){\equiv}{\mathcal{L}}^{-1}[k(s);\nu]$ and
$\widehat{J}(\nu){\equiv}{\mathcal{L}}^{-1}[h(s);\nu]$
respectively. Therefore the solution of the inverse Laplace
transform of coefficients $k(s)$ and $h(s)$ can be converted to
$\nu$-space as
\begin{eqnarray}
\widehat{\eta}(\nu)&=&\frac{32}{3}e^{(-2\nu)}+e^{-\frac{3}{2}\nu}[\frac{64}{21}\sqrt{7}(\ln(2)+3)\sin(\frac{1}{2}\sqrt{7}\nu)\nonumber\\
&&-\frac{64}{3}(\ln(2)+1)\cos(\frac{1}{2}\sqrt{7}\nu)],
\end{eqnarray}
and
\begin{eqnarray}
\widehat{J}(\nu)=e^{-\frac{3}{2}\nu}[4\cos(\frac{1}{2}\sqrt{7}\nu)
-\frac{4}{7}\sqrt{7}\sin(\frac{1}{2}\sqrt{7}\nu)].
\end{eqnarray}
Consequently, the general relation between the structure functions
in $x$-space is given by
\begin{eqnarray}
F_{L}(x,Q^{2})&=&\frac{\alpha_{s}(Q^{2})}{4\pi}\int_{x}^{1}F_{2}(y,Q^{2})\eta(\frac{x}{y})\frac{dy}{y}\nonumber\\
&&+\int_{x}^{1}\frac{{\partial}F_{2}(y,Q^{2})}{\partial{\ln}Q^{2}}J(\frac{x}{y})\frac{dy}{y},
\end{eqnarray}
where
\begin{eqnarray}
\frac{{\partial}F_{2}(x,Q^{2})}{\partial{\ln}Q^{2}}&=&
F_{2}(x,Q^{2})[\frac{{\partial}{\ln}D(Q^{2})}{\partial{\ln}Q^{2}}\nonumber\\
&&+\frac{{\partial}{\ln}\sum_{m=0}^{2}A_{m}(Q^{2})L^{m}}{\partial{\ln}Q^{2}}],\nonumber
\end{eqnarray}
and
\begin{eqnarray}
\eta(\frac{x}{y})&=&
\frac{32}{3}(\frac{x}{y})^{2}+(\frac{x}{y})^{3/2}[\frac{64}{21}\sqrt{7}(\ln(2)+3)\sin(\frac{1}{2}\sqrt{7}{\ln}(\frac{y}{x}))\nonumber\\
&&-\frac{64}{3}(\ln(2)+1)\cos(\frac{1}{2}\sqrt{7}{\ln}(\frac{y}{x}))
],\nonumber\\
J(\frac{x}{y})&=&(\frac{x}{y})^{3/2}
[4\cos(\frac{1}{2}\sqrt{7}{\ln}(\frac{y}{x}))-\frac{4}{7}\sqrt{7}\sin(\frac{1}{2}\sqrt{7}{\ln}(\frac{y}{x}))].\nonumber
\end{eqnarray}
Therefore the longitudinal structure function due to the
Laplace-transform method is obtained by the parametrization of the
structure function  $F_{2}$ and its derivative
${{\partial}F_{2}(x,Q^{2})}/{\partial{\ln}Q^{2}}$. In order to
make the effect of production threshold for charm quark, I use the
rescaling variable $\chi$ which introduced by Aivazis, Collins,
Olness and Tung (ACOT) in Ref.[19]. Therefore, the longitudinal
structure function is defined by the rescaling variable $\chi$
where $ \chi=x(1+\frac{4m_{c}^{2}}{Q^{2}})$. The rescaling
variable $\chi$ at high $Q^{2}$ values ($m_{c}^{2}/Q^{2}{\ll}1$)
reduces to the Bjorken variable $x$ as $\chi{\rightarrow}x$ [19].
The running charm mass is obtained as
$m_{c}=1.29^{+0.077}_{-0.053} \mathrm{GeV}$, where the
uncertainties are obtained through adding the experimental fit,
model and parametrization uncertainties in quadrature [20,21].\\

\subsection{Results and Discussion}

In Ref.[4] the QCD parameter $\Lambda$ has been extracted due to
$\alpha_{s}(M_{z}^{2})=0.1166$, which for four number of active
flavor is defined by $ \Lambda=136.8~ \mathrm{MeV}$. With the
explicit form of the proton structure function, I can proceed to
extract the longitudinal structure function $F_{L}(x,Q^{2})$  from
data mediated by the parametrization of the structure function
$F_{ 2}(x,Q^{2})$ and its derivative. I have calculated the
$x$-dependence of the longitudinal structure function at several
fixed values of $Q^{2}$ corresponding to H1 collaboration data
[21,22]. Results are presented and compared with H1 collaboration
data [21,22] and the parametrization of the longitudinal structure
function  $F_{L}(x,Q^{2})$ [3] at leading-order approximation in
Fig.1. The error bands illustrated in this figure are into the
charm-quark mass uncertainty and the statistical errors in the
parametrization of $F_{2}(x,Q^{2})$, where the fit parameter
errors are shown in Table I. It is seen that, for all values of
the presented $Q^{2}$ with respect to the rescaling variable, the
extracted longitudinal structure function due to the Laplace
transform method is comparable with the H1 collaboration data. In
order to present more detailed discussions on our findings,  the
results for the longitudinal structure function compared with
leading and next-to-leading order (LO and NLO respectively) of
CJ15 [23] in this figure.
 Also the obtained longitudinal structure functions
compared with  the Mellin transforms method [4] at leading order approximation in Fig.1.\\
The effect of the charm quark mass  in the splitting functions
into the rescaling variable is considered in Fig.2. These results
(with the rescaling and Bjorken variables) in Fig.2 compared with
CJ15 at LO and NLO approximations and H1 collaboration data. As
can be seen in Fig.2, the results with Bjorken variable are
comparable with the results in Ref.[4] which is based on the
Mellin transforms method. Indeed, the Mellin and Laplace transform
methods have the same behavior with two difference schemes. These
results compared with the obtained longitudinal structure
functions with respect to the rescaling variable. Indeed, the
rescaling variable improves the longitudinal structure functions
in comparison with the Mellin transform that is based on the
Bjorken
variable when the longitudinal structure functions compared with H1 collaboration data.\\
The accuracy of the Laplace and Mellin transform methods at LO
approximation in comparison with the NLO CJ15 [23] is illustrated
in Fig.3 by the ratio
\begin{eqnarray}
r_{F_{L}}&=&\frac{F_{L}^{\mathrm{Analytic}}(x,Q^{2})}{F_{L}^{\mathrm{CJ15NLO}}(x,Q^{2})},\nonumber
\end{eqnarray}
where analytic return to the Laplace transforms method at LO
approximation by the Bjorken and rescaling variables and also the
Mellin transforms method at LO approximation by the Bjorken
variable. In the left-hand column of Fig.3, I show the fractional
accuracy $r_{F_{L}}$ for Laplace and Mellin transforms methods due
to the Bjorken and rescaling variables in comparison with the NLO
CJ15 at $Q^{2}=45~\mathrm{GeV}^{2}$. Fig.3 shows that the
longitudinal structure functions obtained at LO approximation by
Laplace transforms method from the rescaling variable are
consistent with the H1 collaboration data at
$Q^{2}=45~\mathrm{GeV}^{2}$ and can be extended to all $Q^{2}$
values. Both the Laplace and Mellin transforms method at LO
approximation by the Bjorken variable  are not agreement with the
H1 data region in the domain $x{\simeq}10^{-3}$. Also, in the
right-hand column of Fig.3,  the fractional accuracy $R_{F_{L}}$
at $Q^{2}=5~\mathrm{GeV}^{2}$ for Laplace transforms method due to
the Bjorken and rescaling variables in comparison with the Mellin
transforms method at LO approximation by the Bjorken scaling is
shown by the following form
\begin{eqnarray}
R_{F_{L}}&=&\frac{F_{L}^{\mathrm{Laplace}}(x,Q^{2})}{F_{L}^{\mathrm{Mellin}}(x,Q^{2})}.\nonumber
\end{eqnarray}
This figure shows that $R_{F_{L}}$ for Bjorken scaling is
essentially constant and approximately equal to one for all
$x{\lesssim}0.01$. The discrepancies between the Laplace and
Mellin by the Bjorken scaling  are small, but they are large when
 the rescaling variable is considered. Indeed these results due to
 the rescaling variable are comparable with the CJ15 NLO and H1
 collaboration data in H1 data domain.\\
In Fig.4, I show the $Q^{2}$-dependence of the longitudinal
structure function at low $x$. Results of calculations and
comparison with the H1 collaboration data [21,22] are presented in
this figure (i.e., Fig.4), where the charm quark mass effects is
considered in the rescaling variable. These results have been
performed at fixed value of the invariant mass $W$ as $W=230~
\mathrm{GeV}$. Figure 4 shows that the parametrization of parton
distributions provides correct behaviors of the extracted
$F_{L}(x,Q^{2})$ in comparison with the Mellin transforms method.
Over a wide range of variable $Q^{2}$, the extracted longitudinal
structure functions are in a good agreement with experimental data
in comparison with the parametrization of the longitudinal
structure function $F_{L}(x,Q^{2})$ at leading-order
approximation. At low values of $Q^{2}$, the extracted results are
still above the experimental data. The error bands illustrated in
this figure are into the charm-quark mass uncertainty  and the
statistical errors in the parametrization of  the structure
function $F_{2}(x,Q^{2})$, where the fit
parameter errors are shown in Table I.\\
Also comparison between the longitudinal structure functions, into
the rescaling and Bjorken scaling variables, are shown in Fig.4
for a wide range of $Q^{2}$. These results (with the rescaling and
Bjorken variables) in Fig.4 compared with the H1 collaboration
data [21]. As can be seen in this figure, the Laplace and Mellin
transform methods in the Bjorken scaling are comparable together
in a wide range of $Q^{2}$ values at fixed center-of-mass energy.
At fixed value of the invariant mass $W=230~\mathrm{GeV}$, the
rescaling variable improve the longitudinal structure function
results in comparable with the H1 collaboration data.\\

In conclusion, I have presented a certain theoretical model to
describe the longitudinal structure function based on the Laplace
transform method  at low values of $x$. A detailed analysis has
been performed to find an analytical solution of the longitudinal
structure function from the parametrization of the structure
function  $F_{2}(x,Q^{2})$ and its derivative. The effect of
massive quarks in the splitting functions is considered into the
rescaling variable. The calculations are consistent with the H1
collaboration data from HERA collider. As a next step, I plan to
take into account the high-order corrections  in a similar manner
as in Ref. [4] (Phys.Rev.D {\bf99}, 096019 (2019)) since these
corrections are important in the region of small $Q^{2}$, also the
non-linear corrections improve the longitudinal structure function
behavior
 at low $Q^{2}$.\\

\subsection{ACKNOWLEDGMENTS}

The author is thankful to the Razi University for financial
support of this project.\\

\subsection{Appendix A}
The proton structure function parameterized in Ref.[5] provide
good fits to the HERA data at low $x$ and large $Q^{2}$ values.
The explicit expression for the proton structure function, with
respect to the Block-Halzen fit, in a range of the kinematical
variables $x$ and $Q^{2}$, $x{\leq}0.1$ and
$0.15~\mathrm{GeV}^{2}<Q^{2}<3000~\mathrm{GeV}^{2}$, is defined by
the following form
\begin{eqnarray}
F^{\gamma p}_{ 2}(x,Q^{2})& =& D(Q^{2})(1-
x)^{n}[C(Q^{2})+A(Q^{2})\ln(\frac{1}{x}\frac{Q^{2}}{Q^{2}+\mu^{2}})\nonumber\\
&&+B(Q^{2})\ln^{2}(\frac{1}{x}\frac{Q^{2}}{Q^{2}+\mu^{2}})],
\end{eqnarray}
where
\begin{eqnarray}
 A(Q^{2})& =& a_{0} + a_{1} {\ln}(1+\frac{Q^{2}}{\mu^{2}}) + a_{2}{\ln}^{2}(1+\frac{Q^{2}}{\mu^{2}})
 ,\nonumber\\
B(Q^{2})& =& b_{0} + b_{1} {\ln}(1+\frac{Q^{2}}{\mu^{2}}) +
b_{2}{\ln}^{2}(1+\frac{Q^{2}}{\mu^{2}})
 ,\nonumber\\
C(Q^{2})& =& c_{0} + c_{1}
{\ln}(1+\frac{Q^{2}}{\mu^{2}}),\nonumber\\
D(Q^{2})& =& \frac{Q^{2}(Q^{2}+\lambda M^{2})}{(Q^{2}+M^{2})^2}.
\end{eqnarray}
Here $M$ is the effective mass and $\mu^{2}$ is a scale factor.
The additional parameters with their statistical errors are given
in Table I.\\
\begin{table}[h]
\caption{ The effective parameters at low $x$ for
$0.15~\mathrm{GeV}^{2}<Q^{2}<3000~\mathrm{GeV}^{2}$ provided by
the following values. The fixed  parameters are defined by the
Block-Halzen fit to the real photon-proton cross section as
$M^{2}=0.753 \pm 0.068~ \mathrm{GeV}^{2}$, $\mu^2 = 2.82 \pm
0.290~ \mathrm{GeV}^{2}$ and $c_{0} = 0.255 \pm 0.016$ [5].}
\begin{tabular} {cccc}
\toprule \\  \multicolumn{2}{c}{parameters \quad \quad \quad ~~~~~~~~~~~~~~~~value}    \\ &&&\\ \hline \\ &&&\\
  $a_{0} $  &   \quad  $8.205\times 10^{-4}~~  \pm  4.62\times10^{-4} $  \\

  $a_{1} $  &   \quad   $-5.148\times 10^{-2}\pm 8.19\times10^{-3}$  \\

  $a_{2}$   &    \quad  $-4.725\times 10^{-3}\pm 1.01\times10^{-3}$   \\  &&&\\

 $b_{0}$   &   \quad   $2.217\times 10^{-3}\pm 1.42\times10^{-4} $ \\

 $b_{1}$   &   \quad   $1.244\times 10^{-2}\pm 8.56\times10^{-4}$  \\

 $b_{2}$    &    \quad  $5.958\times 10^{-4}\pm 2.32\times10^{-4} $ \\ &&& \\

$c_{1}$& \quad  $1.475\times 10^{-1}~\pm 3.025\times10^{-2}$ & &\\

$n$& \quad  $11.49\pm 0.99$ & &\\

$\lambda$& \quad  $2.430~\pm 0.153$ & &\\

$\chi^{2}(\mathrm{goodness~ of~ fit})$ &  \quad  $0.95$ & &\\
\hline

\end{tabular}
\end{table}

\newpage{
\section{References}

1. J.Abelleira Fernandez et al., [LHeC Collaboration], J.Phys.G39,
075001(2012).\\
2. P. Agostini et al. (LHeC Collab- oration and FCC-he Study
Group), CERN-ACC-Note- 2020-0002,
arXiv:2007.14491 [hep-ex] (2020).\\
3. G.Altarelli and G.Martinelli, Phys.Lett.B\textbf{76}, 89(1978).\\
4. L.P.Kaptari et al., Phys.Rev.D {\bf99}, 096019 (2019); JETP Lett.{\bf 109}, 281(2019).\\
5. M. M. Block, L. Durand and P. Ha, Phys. Rev. D {\bf89}, 094027
(2014).\\
6. G.R.Boroun, JETP lett.{\bf114}, 3(2021); Eur.Phys.J.Plus
{\bf135}, 68(2020);
Phys.Rev.C {\bf97}, 015206(2018); Eur.Phys.J.Plus {\bf129}, 19(2014).\\
7. J.Blumlein, J.Phys.G {\bf19}, 1623 (1993); A.M.Cooper-Sarkar,
Z.Phys.C {\bf39}, 281 (1988);  J.Schwartz, arXiv [hep-ex]:1010.1023 (2010); S.Simula, Phys.Lett.B {\bf574}, 189(2003).\\
8. B.Rezaei and G.R.Boroun, Eur.Phys.J.A {\bf56}, 262(2020).\\
9. N.Baruah, M.K.Das and J.K.Sarma, Eur.Phys.J.Plus {\bf129}, 229
(2014).\\
10. G.R.Boroun, B.Rezaei and J.K.Sarma, Int.J.Mod.Phys.A {\bf29},
1450189 (2014).\\
11. A.V.Kotikov, A.V.Lipatov and N.P.Zotov, J.Exp.Theo.Phys.
{\bf101}, 811 (2005); Eur.Phys.J.C {\bf27}, 219(2003); A.V.Kotikov
and G.Parente, Mod.Phys.Lett.A {\bf12}, 963 (1997); A.V.Kotikov, J.Exp.Theo.Phys. {\bf80}, 979 (1995).\\
12. G.R.Boroun and B.Rezaei, Eur.Phys.J.C {\bf72}, 2221 (2012); Phys.Lett.B {\bf816}, 136274 (2021).\\
13. M.M.Block, L.Durand, P.Ha and D.W.Mckay, Phys.Rev.D {\bf83}, 054009 (2011).\\
14. G.R.Boroun, S.Zarrin and F.Teimoury, Eur.Phys.J.Plus {\bf130},
214 (2015).\\
15. H.Khanpour, A.Mirjalili and S.Atashbar Tehrani, Phys.Rev.C
{\bf95}, 035201 (2017).\\
16. G.R.Boroun and B.Rezaei, Eur.Phys.J.C {\bf73}, 2412 (2013).\\
17. S.M.Moosavi Nejad, H.Khanpour, S.Atashbar Tehrani and
M.Mahdavi, Phys.Rev.C
{\bf94}, 045201 (2016).\\
18. S.Dadfar and S.Zarrin, Eur.Phys.J.C {\bf80}, 319 (2020).\\
19. M.A.G.Aivazis et al., Phys.Rev.D {\bf50}, 3102 (1994).\\
20. H1 and ZEUS Collaborations (Abramowicz H. et al.), Eur. Phys.
J. C {\bf78}, 473 (2018).\\
21. H1 Collab. (V.Andreev , A.Baghdasaryan, S.Baghdasaryan et
al.), Eur.Phys.J.C{\bf74},
2814(2014).\\
22. H1 Collab. (F.D.Aaron , C.Alexa, V.Andreev et al.),Eur.Phys.J.C{\bf71}, 1579 (2011).\\
23. A.Accardi, L.T.Brady, W.Melnitchouk, J.F.Owens, N.Sato, Phys.Rev.D {\bf93}, 114017 (2016).\\
}

\begin{figure}
\includegraphics[width=1\textwidth]{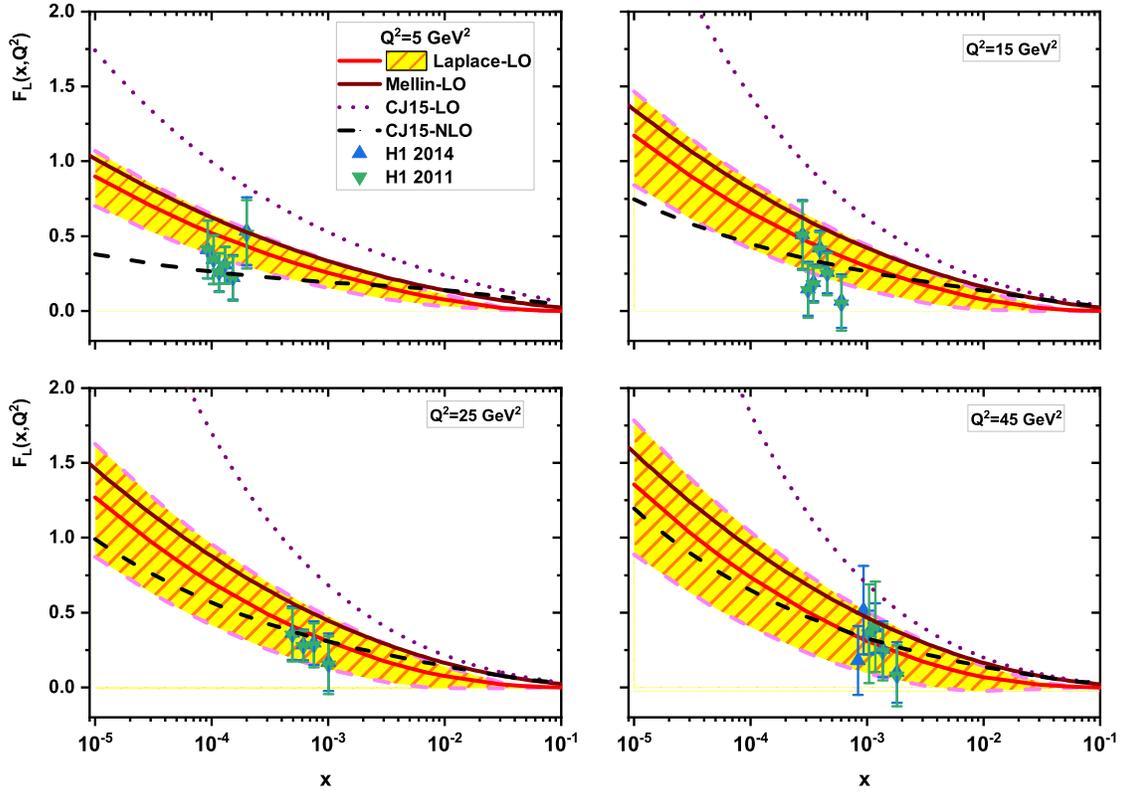}
\caption{The longitudinal structure function, based on the
rescaling variable by the Laplace transform method,  extracted in
comparison with the H1 experimental data (up-triangle H1 2014,
down-triangle H1 2011) [21,22] as accompanied with total errors.
The error bands are due to the charm-quark mass uncertainty  and
the statistical errors in the parametrization of $F_{2}(x,Q^{2})$
and its derivative. The dashed, dot and solid lines represent the
CJ15 at NLO and LO approximations [23] and the Mellin transforms
method [4] at the LO approximation respectively.}\label{Fig1}
\end{figure}
\begin{figure}
\includegraphics[width=1\textwidth]{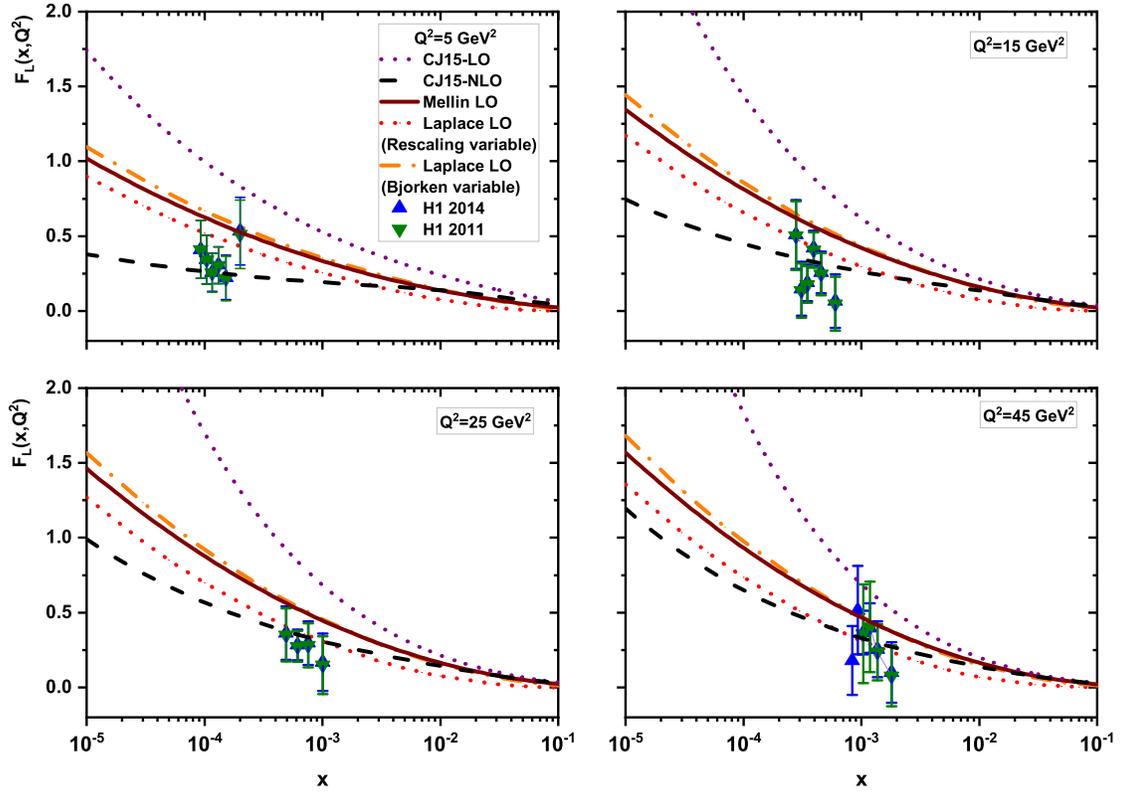}
\caption{The longitudinal structure functions based on the
rescaling (dot lines) and Bjorken (dashed-dot lines) variables
compared with the H1 experimental data (up-triangle H1 2014,
down-triangle H1 2011) [21,22] as accompanied with total errors,
 the CJ15 at NLO and LO approximations  [23] (dashed and dot lines) and the Mellin
transform method [4] (solid lines)  at LO approximation
respectively.}\label{Fig2}
\end{figure}
\begin{figure}
\includegraphics[width=0.55\textwidth]{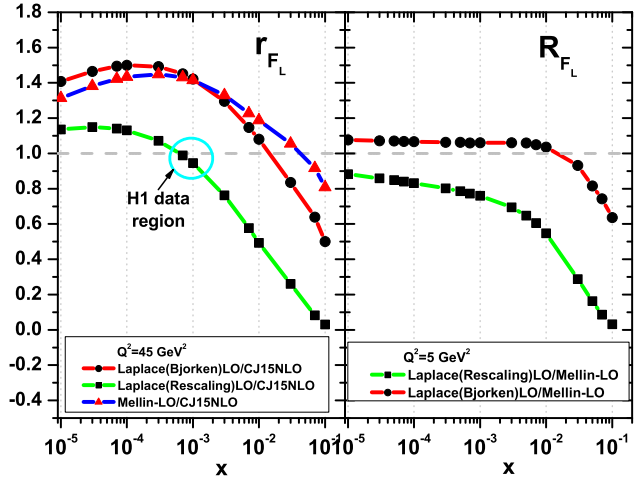}
\caption{Left-hand column: comparison of the $r_{F_{L}}$ obtained
by the Laplace and Mellin transforms methods by the Bjorken and
rescaling variables with CJ15 NLO at $Q^{2}=45~\mathrm{GeV}^{2}$.
Right-hand column: comparison of the $R_{F_{L}}$ obtained by the
Laplace  transforms method by the Bjorken and rescaling variables
with and the Mellin transforms method at
$Q^{2}=5~\mathrm{GeV}^{2}$. }\label{Fig3}
\end{figure}
\begin{figure}
\includegraphics[width=1\textwidth]{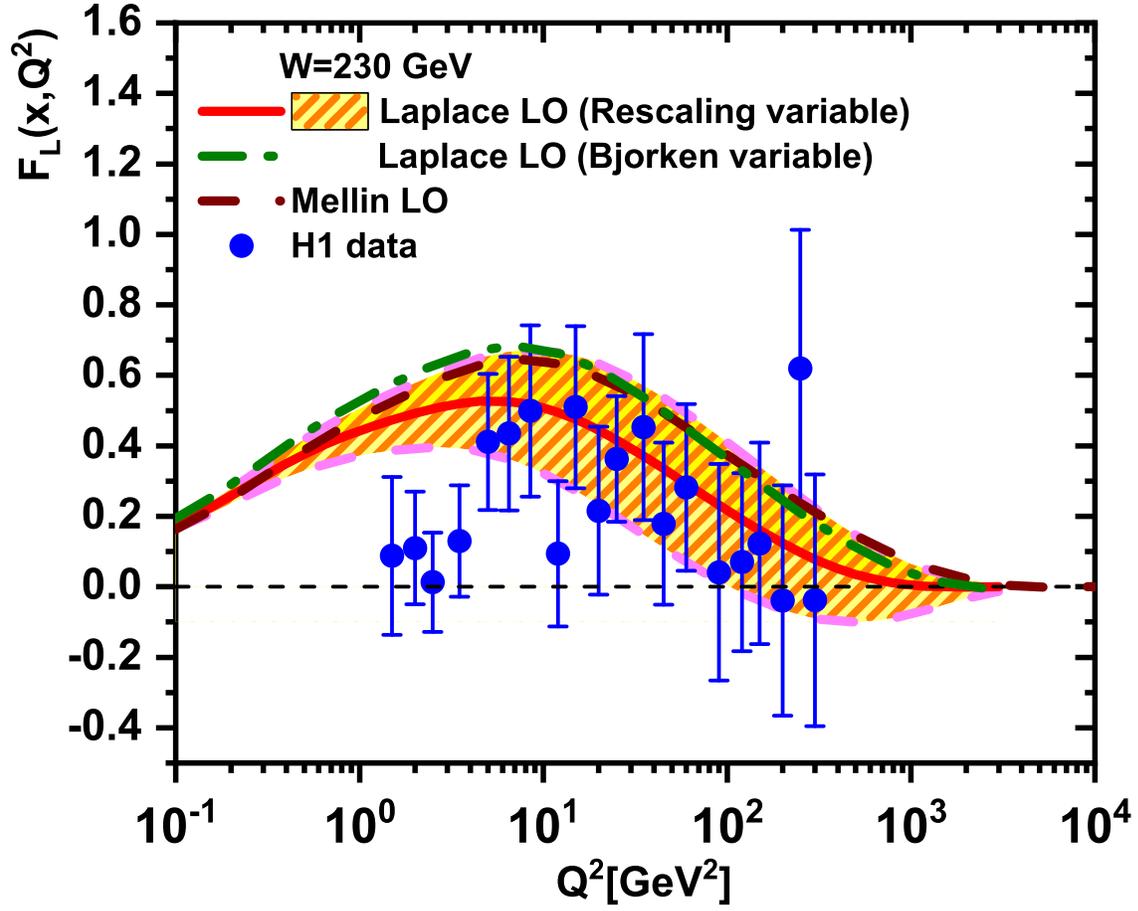}
\caption{$Q^{2}$ dependence of the extracted longitudinal
structure function  at fixed value of the invariant mass $W=230~
\mathrm{GeV}$, into the rescaling (solid curve) and Bjorken
(dashed-dot curve) variables compared with the Mellin transform
method [4](dashed curve) at the LO approximation. The error bands
are due to the charm-quark mass uncertainty  and the statistical
errors in the parametrization of $F_{2}(x,Q^{2})$ and its
derivative. Experimental data by the H1 Collaboration are taken
from Ref. [21] as accompanied with total errors.}\label{Fig4}
\end{figure}

\end{document}